\documentclass[12pt,preprint]{aastex}
\usepackage{amsmath}

\newcommand{\Swift}{{\it Swift}}
\newcommand{\Suzaku}{{\it Suzaku}}
\newcommand{\Swifts}{{\it Swift }}
\newcommand{\Suzakus}{{\it Suzaku }}
\newcommand{\fermi}{{\it Fermi }}
\newcommand{\amati}{${E^{\rm src}_{\rm peak} - E_{\rm iso}}$ }
\newcommand{\ghil}{${E^{\rm src}_{\rm peak}-E_{\rm \gamma}}$ }

\slugcomment{ApJ (accepted)}
\shorttitle{Synchrotron Self-inverse Compton Radiation From Reverse-shock on GRB120326A}
\shortauthors{Urata  et al.}

\begin{document}

\title{Synchrotron self-inverse Compton radiation from reverse shock on GRB120326A}

\author{
Yuji~\textsc{Urata}\altaffilmark{1}, 
Kuiyun~\textsc{Huang}\altaffilmark{2,3},
Satoko~\textsc{Takahashi}\altaffilmark{2,4,5},
Myungshin~\textsc{Im}\altaffilmark{6},
Kazutaka~\textsc{Yamaoka}\altaffilmark{7,8},
Makoto~\textsc{Tashiro}\altaffilmark{9},
Jae-Woo~\textsc{Kim}\altaffilmark{6},
Minsung~\textsc{Jang}\altaffilmark{6},
and 
Soojong~\textsc{Pak}\altaffilmark{10}
%et al.
%%%%
%Masanori~\textsc{Ohno}\altaffilmark{8},
%and 
%Glen~\textsc{Petitpas}\altaffilmark{3}
}

\altaffiltext{1}{Institute of Astronomy, National Central University, Chung-Li 32054, Taiwan, urata@astro.ncu.edu.tw}
\altaffiltext{2}{Academia Sinica Institute of Astronomy and Astrophysics, Taipei 106, Taiwan}
\altaffiltext{3}{Department of Mathematics and Science, National Taiwan Normal University, Lin-kou District, New Taipei City 24449, Taiwan} 
\altaffiltext{4}{Joint ALMA Observatory, Alonso de Cordova 3108, Vitacura, Santiago, Chile}
\altaffiltext{5}{National Astronomical Observatory of Japan, 2-21-1 Osawa, Mitaka, Tokyo 181-8588, Japan}
\altaffiltext{6}{Center for the Exploration of the Origin of the Universe, Department of Physics \& Astronomy, FPRD, Seoul National University, Shillim-dong, San 56-1, Kwanak-gu, Seoul, Korea}
\altaffiltext{7}{Solar-Terrestrial Environment Laboratory, Nagoya University, Furo-cho, Chikusa-ku, Nagoya, Aichi 464-8601, Japan}
\altaffiltext{8}{Division of Particle and Astrophysical Science, Graduate School of Science, Nagoya University, Furo-cho, Chikusa-ku, Nagoya, Aichi 464-8601, Japan}
\altaffiltext{9}{Department of Physics, Saitama University, Shimo-Okubo, Saitama, 338-8570, Japan}
\altaffiltext{10}{School of Space Research, Kyung Hee University, Yongin, Gyeonggi 446-701, Korea}

\begin{abstract}

  We present multi-wavelength observations of a typical long duration
  GRB 120326A at $z=1.798$, including rapid observations using a
  submillimeter array (SMA), and a comprehensive monitoring in X-ray and
  optical. The SMA observation provided the fastest detection to date
    among seven submillimeter afterglows at 230 GHz.
  The prompt spectral analysis, using {\it Swift} and {\it Suzaku}
  yielded a spectral peak energy of $E^{\rm src}_{\rm
    peak}=107.8^{+15.3}_{-15.3}$ keV and equivalent isotropic energy of
  $E_{\rm iso}$ as $3.18^{+0.40}_{-0.32}\times 10^{52}$ erg.
  The temporal evolution and spectral properties in the optical were
  consistent with the standard forward shock synchrotron with jet
  collimation ($6^{\circ}.69\pm0^{\circ}.16$).
  The forward shock modeling using a 2D relativistic hydrodynamic jet
  simulation also determined the reasonable burst explosion and the
  synchrotron radiation parameters for the optical afterglow.  The
  X-ray light curve showed no apparent jet break and the temporal decay index
  relation between the X-ray and optical ($\alpha{\rm
    o}-\alpha_{X}=-1.45\pm0.10$) indicated different radiation
  processes in the X-ray and optical. Introducing synchrotron
  self-inverse Compton radiation from reverse shock is a possible
  solution, and a the detection and the slow decay of the afterglow in
  submillimeter supports that this is a plausible idea.  The observed
  temporal evolution and spectral properties as well as forward shock
  modeling parameters, enabled to determine reasonable functions to
  describe the afterglow properties. Because half of events share
  similar properties in the X-ray and optical to the current event,
  GRB120326A will be a benchmarks with further rapid follow-ups, using
  submillimeter instruments such as SMA and ALMA.

\end{abstract} 


\section{Introduction}

Gamma-ray bursts (GRBs) are among the most powerful explosions in the
universe and are observationally characterized with intense
short flashes primarily in a high-energy band (so-called “prompt emission”)
and long-lived afterglows observed from X-ray to radio wavelength. 
The GRB afterglow is believed to involve a relativistically expanding
fireball \citep[e.g.,][]{fireball}. Interstellar matter (ISM)
influences the fireball shell after it has been collected and
considerable energy is transferred from the shell to the ISM. The
energy transfer is caused by two shocks: a forward shock propagating
into the ISM and a reverse shock propagating into the shell.  It is
also believed that the forward shock produces long-lived afterglows
and the reverse shock generates short-lived bright optical flashes\citep[e.g.,][]{990123opt}
and/or intense radio afterglows\citep[e.g.,][]{990123radio}.

A number of afterglows have been monitored densely in X-ray and
optical bands since launch of the {\it Swift} satellite
\citep{swiftsat}, and significant number of afterglows showed the
different temporal evolutions in X-ray and optical bands. These
results indicated that the simple forward shock model cannot explain
their behavior altogether, and additional processes
are required \citep[e.g.,][]{pa2006,050319,normal,li12}.
Inverse-Compton scattering and/or reverse shocks may play the
important role to solve the problem.  \citet{IC} suggested the local
inverse-Compton scattering to describe the X-ray faster decay comparing
with that of optical. \citet{ssc} introduced a synchrotron self-inverse
Compton radiation from a reverse shock to explain X-ray flare and its
early afterglows.
Thus, confirming the existence of reverse shocks at particularly
longer wavelengths and ascertaining their typical occurrence
conditions is critical. 
Because the expected lifetime of reverse shock at longer wavelengths
is substantially longer than those at optical wavelengths, decoding
radiations into forward and reverse shock components is possible.  In
addition, numerous rapid optical follow-ups are missing the reverse
shock components; although, several successful detections at optical
wavelengths have been made.

The possible reason of the missing reverse shock component would be
that typical reverse shock synchrotron frequency is much below the
optical band. Submillimeter (submm) observations are the key element
to catch reverse shock and to understand emission mechanism of GRB
afterglows. Searching for reverse shock emission in the submm
wavelength would test this possibility.  These submm observations also
provide clean measurements of source intensity, unaffected by
scintillation and extinction.
However, no systematic submm observational studies in the early
afterglow phase exist. This has remained so in reverse shock studies
for some time. One of the main reasons the absence of dedicated submm
telescopes and strategic follow-ups with rapid response that involve
employing open-use telescopes for this challenging observations. In
addition, it is nearly impossible to have rapid (several hrs after the
burst) follow-ups with current open-use telescopes that require manual
preparation of the observational scripts. In addition to this
technical problem, sensitivities of current submm telescopes except
ALMA are not good enough to detect number of afterglows in the submm
band \citep[e.g.,][]{postigo}. Hence, rapid careful target selections
are required to conduct effective submm follow-up observations
using open-use resources.

GRB\,120326A was detected and localized using \Swift
\citep{swift}. The \Suzaku/Wide-band All-Sky Monitor (\Suzaku/WAM) and
\fermi Gamma-ray Burst Monitor (\fermi/GBM) also detected this burst,
and quick look spectral analysis were reported
\citep{wam-obs,gbm-obs}.
%m
The optical afterglow was discovered by \citet{ot} and observed at the
early stage by using several telescopes. The optical afterglow also
exhibited remarkable rebrightening \citep{ot-reb}.
The afterglow at submm and radio bands was also detected using SMA
\citep[SMA;][]{smafollow}, Combined Array for Research in
Millimeter-wave Astronomy \citep[CARMA;][]{carma}, and Expanded Very
Large Array \citep[EVLA;][]{evla}. The SMA observation provided the
fastest afterglow detection (about $4.6\times10^{4}$ s after the
burst) among seven submm afterglows at 230 GHz which are mostly
detected about $1\times10^{5}$ s ($\sim$1 d) after the
bursts. Although numerous follow-ups in various wavelength have been
conducted, the submm afterglow monitoring from earlier phase
($<1\times10^{5}$ s) is still rare, and essential observational
approach to understand the puzzle of afterglow radiation.
The redshift was determined to be $z$=1.798, according to a series of
metal absorption features \citep{redshift}. We used $f(t,\nu)\propto
t^{\alpha}\nu^{\beta}$ to express the afterglow properties.

%Standard afterglow model.\\
%Submm observations.
%This event.\\

\section{Observations}

%%% Prompt observations with Swift 
The \Swifts Burst Alert Telescope (\Swifts/BAT) triggered and located
GRB\,120326A at 01:20:29 (T$_{0}$) UT on March 26, 2012. \Swifts immediately slewed
to the burst, and the XRT initiated follow-up observations
at 59.5 s after the burst. The X-ray afterglow was identified and
localized at RA$=18^{\rm h}15^{\rm m}36^{\rm s}.47$,
Dec$=+69^{\circ}15'37".0$, with an uncertainty of $4".1$. The X-ray
afterglow was observed using the XRT until $\sim5\times10^{5}$
s.
UVOT also obtained images by using the White filter starting 67 s after
the burst and no counterpart in the band was observed.

\begin{figure}
\epsscale{.90}
\plotone{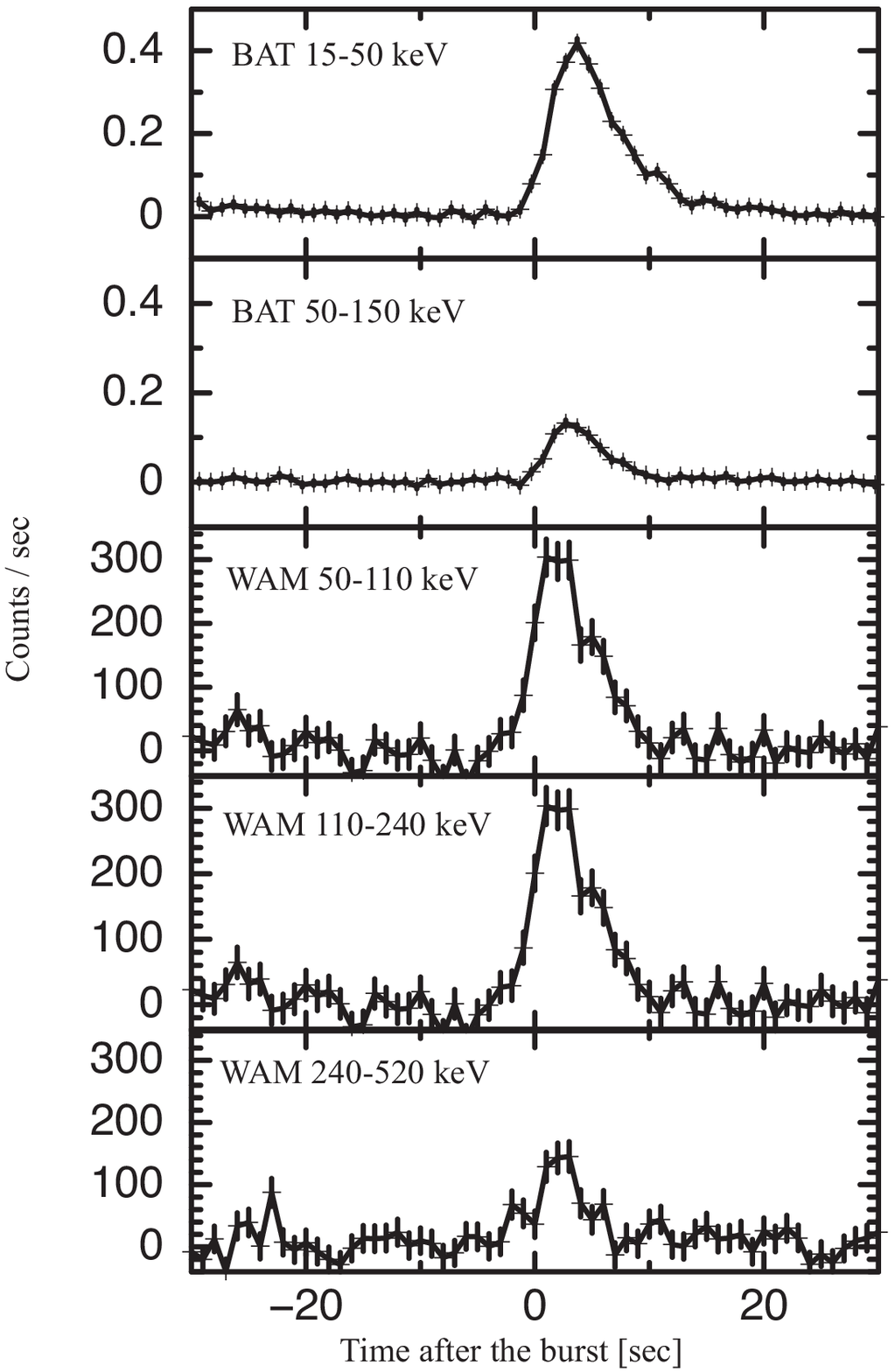}
\caption{Prompt $\gamma$-ray light curves observed by \Swift/BAT and \Suzaku/WAM. The trigger time of \Swift/BAT is used as T$_{0}$.\label{prompt}}
\end{figure}

%%% Prompt observations with Suzaku
The {\it Suzaku}/WAM also triggered the burst at 01:20:31.9
(T$_{0}+2.9$ s) UT on March 26, 2012. The WAM \citep{wam} is a lateral
shield of the Hard X-ray Detector \citep{hxd} on board the \Suzakus
 satellite \citep{suzaku} and is a powerful GRB spectrometer covering
an energy range of 50$-$5000 keV to determine prompt spectral energy
peaks $E_{peak}$ \citep[e.g.,][]{ohno08, tashiro, 071010b}.  As shown
in Figure \ref{prompt}, the prompt X-ray and $\gamma$-ray light curves
observed using the \Swift/BAT and \Suzaku/WAM exhibited a single
fast-rise-exponential-decay (FRED) structure.

%%%% Optical observations with Lulin, LOAO, CQENAM?
We performed the optical afterglow observations using the Lulin
One-meter Telescope \citep{lot} and the LOAO robotic 1-m telescope
\citep{loao,lee} within the framework of the EAFON \citep{eafon}.  Four
color observations were made with LOT on the night of March 26. The
LOAO data were obtained in $R$ band filter from 2013 March 26 to April 2.
The afterglow was also observed with Camera for Quasars in Early
Universe \citep[Camera for QUasars in the EArly
uNiverse;][]{cquean,cquean2, cquean3} on the 2.1-m Otto-Struve
telescope of the McDonald observatory, Texas, USA. The data were
obtained in $g, r, i, z$, and $Y$ filters, starting at 2013 March 26,
10:09:22 (UT), and continued till April 2.  
The logs for both observations are summarized in Table \ref{tbl-1}.

\begin{figure}
\epsscale{1.0}
\plotone{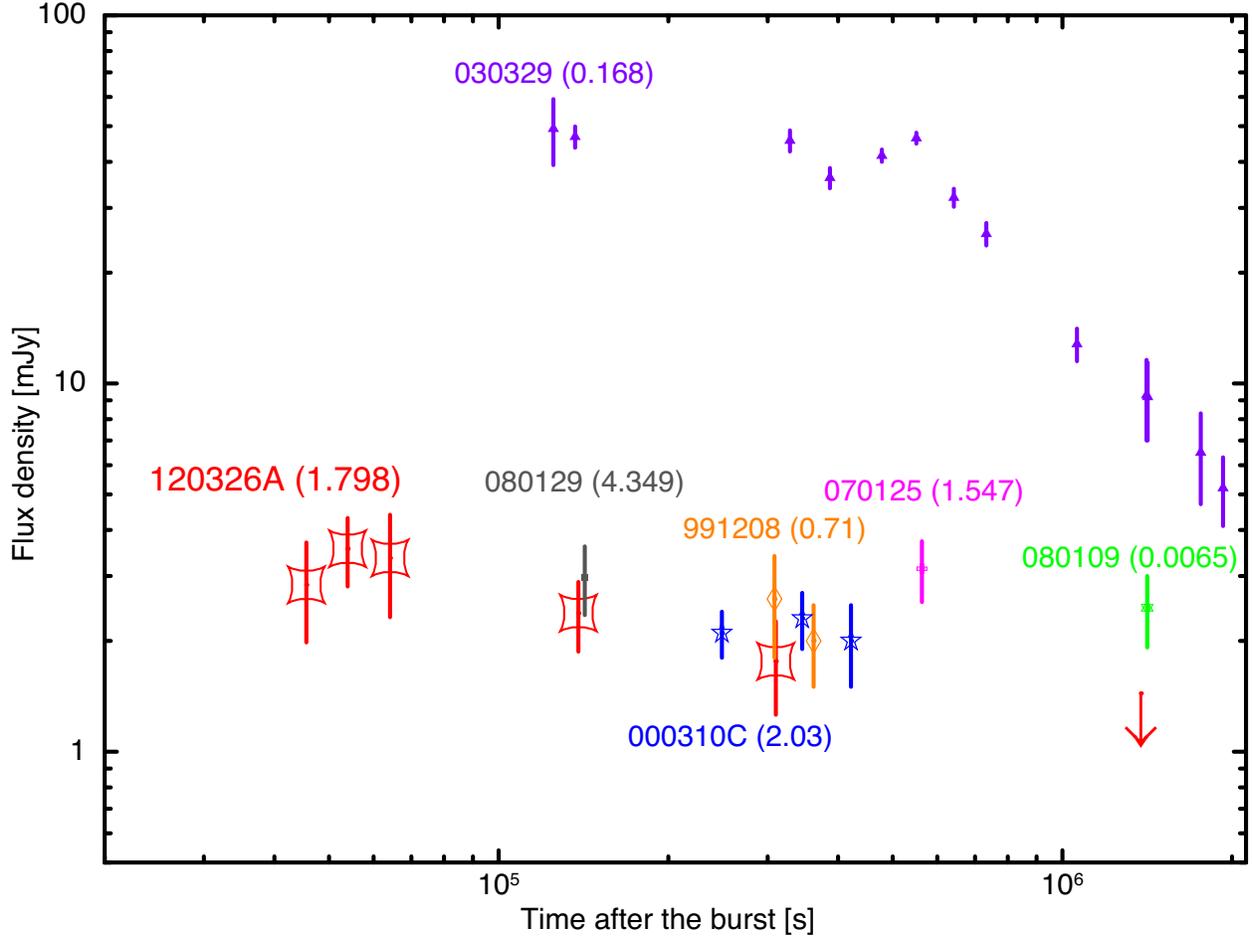}
\caption{Light curves of all the submm afterglows detected in 230 GHz band to date. The SMA observation on GRB120326A provides the earliest submm detection and continuous monitoring. The values noted in brackets is redshift of each event. The samples are taken from \citet{991208} for 991208, \citet{000301c} for 0000301C, \citet{030329a, 030329b} for 030329, \citet{070125} for 070125, \citet{080109} for 080109, and \citet{080129} for 080129.\label{submm}}
\end{figure}

%%%% SMA
We also triggered the submm continuum follow-up observations by using
the seven 6-m antennas of SMA \citep{sma}. The first continuum
observation at 230 GHz (with an 8 GHz bandwidth) was conducted at
10:15:05 on March 26, 2012, about $4.6\times10^{4}$ s after the BAT
trigger. As \citet{smafollow} reported, the submm counterpart was
observed at the location of the X-ray and optical afterglow. The
continuous monitoring using the SMA was also performed at the same
frequency setting on March 27, 29, 31, and April 6, and 11. Table
\ref{tbl-2} summarizes the scientific observations that were conducted
for 4 nights, because of weather conditions and antenna
reconfiguration.
Figure \ref{submm} shows submm light curves of all the GRB afterglows
detected at the 230 GHz to date.  Among all seven events, we
successfully detected the earliest submm afterglow on GRB120326A.
A possible reason for this successful submm monitoring was the target
selection using the quick optical follow-ups.

\section{Analysis and Results}

\begin{figure}
\epsscale{.90}
\plotone{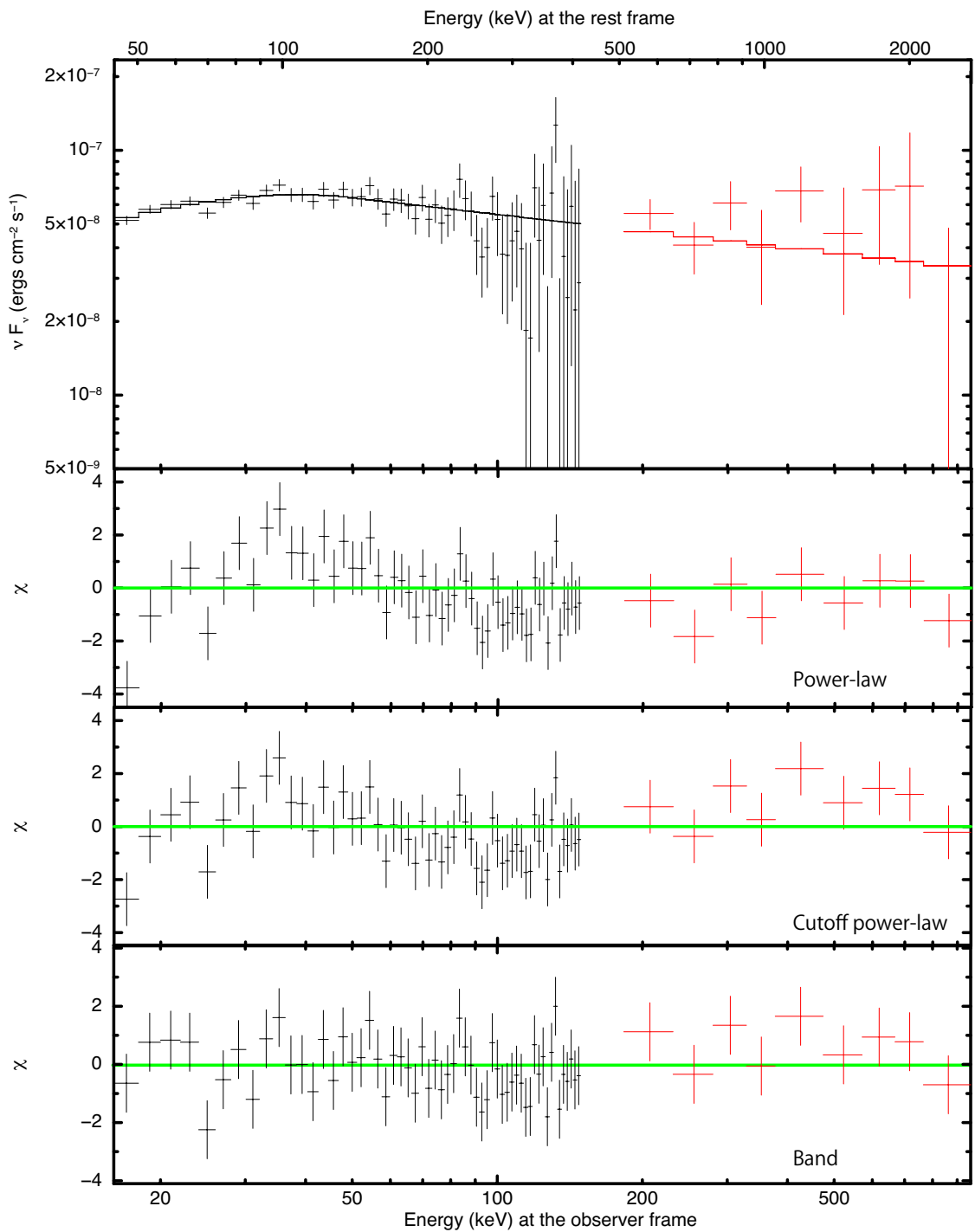}
\caption{Time-averaged spectrum of the prompt emission, observed using \Swift/BAT and
  \Suzaku/WAM. The BAT and WAM data are shown in black and red colros,
  respectively. In the top panel, the solid lines indicate the
  best-fit Band function model. The lower panels show residuals for
  fitting with power-law (2nd), power-law with exponential cutoff
  (3rd), and Band (4th) models.\label{prompt-spec}}
\end{figure}

%%%%%% Joint Prompt spectrum fitting %%%%%%%%
The BAT data were analyzed using the standard BAT analysis software
included in HEADAS v6.12.  The time averaged spectrum (15 -- 150 keV)
from T$_{0}-2$ to T$_{0}+11$ s was extracted using {\tt
  batgrbproduct}.  Response matrices were generated by the task {\tt
  batdrmgen}, using the latest spectral redistribution matrices.
The WAM spectral and temporal data were extracted using {\tt hxdmkwamlc} and
{\tt hxdmkwamspec} in HEADAS version 6.12. The background was estimated
using the fitting model described in \citet{sugita}.  Response
matrices were generated by the WAM response generator as described in
\citet{ohno08}.
We used three models for the joint spectral fitting: power-law,
power-law with exponential cutoff, and Band function model.  
As shown in Figure \ref{prompt-spec}, the spectrum was reasonably
fitted with the Band function. The fitting yielded a low-energy photon
index of $1.17^{+0.53}_{-0.32}$, high-energy photon index of
$2.23^{+0.09}_{-0.11}$ and $\nu F \nu$ spectrum peak energy in the
source frame $E^{\rm src}_{\rm peak}$ of $107.8^{+15.2}_{-15.3}$ keV
($\chi^{2}/\nu=0.92$ for $\nu=63$).
Both power-law ($\chi^{2}/\nu=1.55$ for $\nu=65$) and power-law with
exponential cutoff ($\chi^{2}/\nu=1.38$ for $\nu=67$) were not
acceptable, leaving a curvature of the residuals around 30-40 keV at
the observer frame (the 2nd and 3rd panele in Figure
\ref{prompt-spec}).
%
%The result is consistent with the quick look result of \fermi/GBM \citep{}.
%
We also estimated the equivalent isotropic radiated energy in the
prompt phase at the 1-10000keV band $\rm E_{\rm iso}$ as
$3.18^{+0.40}_{-0.32}\times 10^{52}$ erg, assuming cosmological
parameters: $H_0=71 {\rm km/s/Mpc}$, $\Omega_m=0.27$, and
$\Omega_{\Lambda}=0.73$.

\begin{figure}
\epsscale{.90}
\plotone{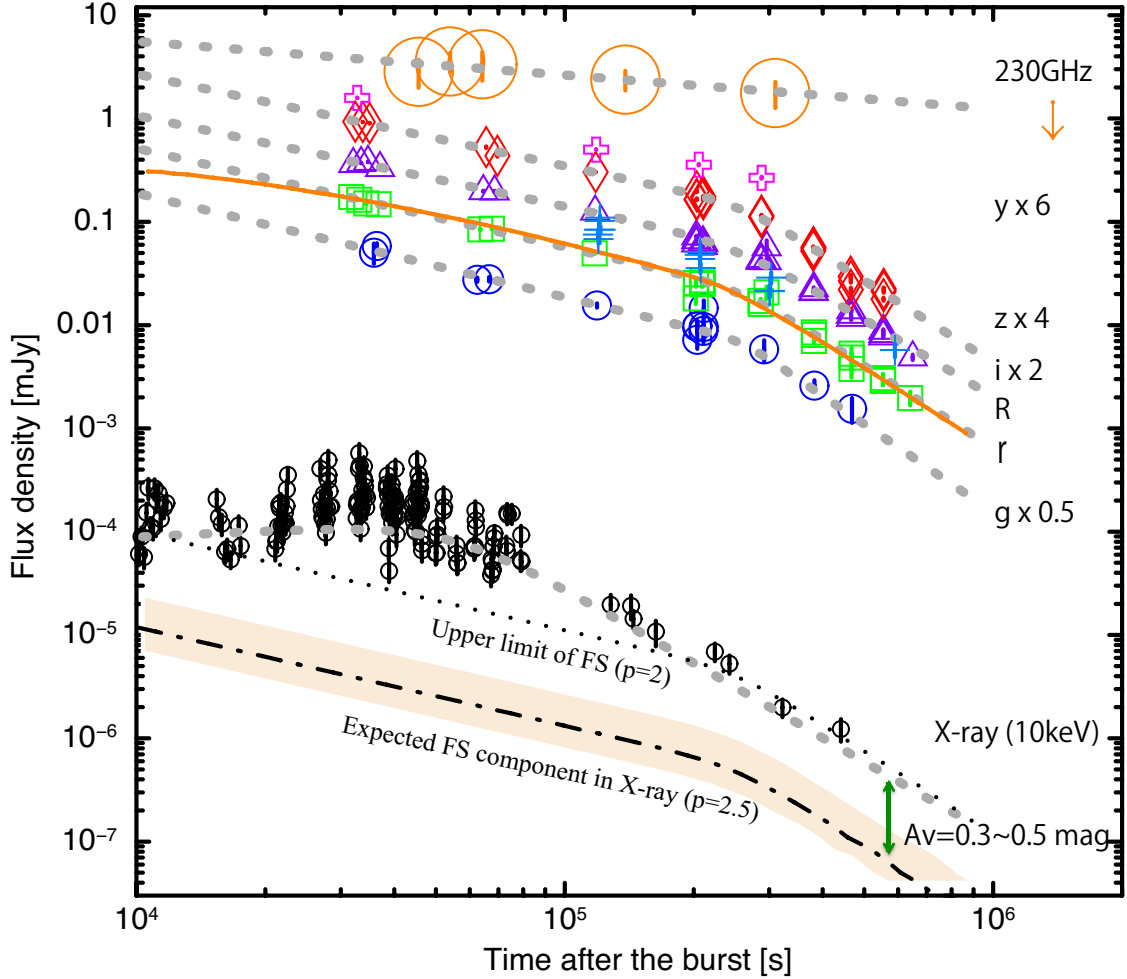}
\caption{X-ray, optical and submm light curves of the GRB120326A afterglow. The grey dotted lines show the best analytical fitted functions described in the text. The orange solid line shows the best modeling function for the $r$-band light curve obtained with the numerical simulation using boxfit. The black dash-dotted and dotted lines indicat the shifted optical light curve to the X-ray bands by factor of $(\nu_{X}/\nu_{opt})^{-p/2}$ with $p=2.5$ and $p=2$, respectively.\label{lc}}
\end{figure}

%%%%% X-ray %%%%%%%
We obtained a reduced \Swift/XRT light curve with a flux density unit at
10 keV from the U.K. Swift Science Data Center \citep{xrt2,xrt1}.  The
light curves shown in Figure \ref{lc} were suitably fitted using the broken power-law model described in
\citep{071010b}, using the best fitted parameters of
$\alpha_{X1}=0.19\pm0.09$, $\alpha_{X2}=-2.35\pm0.15$, and
$t_{bX}=(5.29\pm0.32)\times10^{4}$ s. We also
generated the time-averaged spectrum at a mean time of
$6.47\times10^{4}$ s (from 55741 to 73639 s). The spectra were suitably 
fitted using power-law modified by photo-electric absorptions (galactic and intrinsic), and the
photon index was estimated as $1.95^{+0.18}_{-0.17}$.

\begin{figure}
\epsscale{1.0}
\plotone{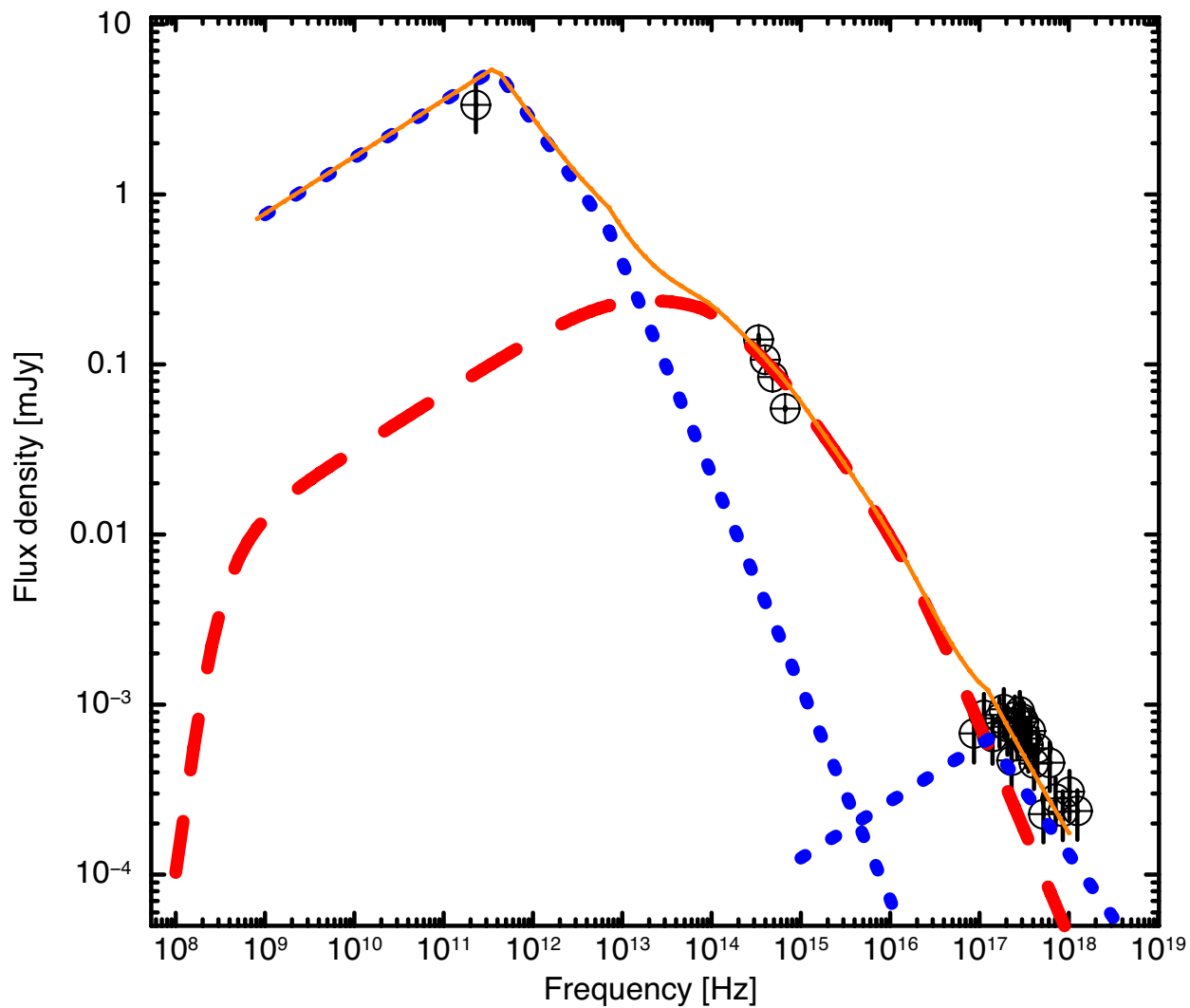}
\caption{The spectral energy distribution at $6.42\times10^{4}$ s after the burst. The red dashed line shows the forward shock synchrotron model spectrum calculated using the boxfit code with the same parameters for the best modeling light curve shown in left panel. The blue dotted lines show the reverse shock synchrotron radiation and its self-inverse Compton component calculated based on \citet{ssc} using the observed values and model function for the forward shock component.\label{lcsed}}
\end{figure}

%%%%%%%% Optical analysis %%%%%%%%
A standard routine including bias subtraction and flat-fielding
corrections was employed to processes the optical data by using the IRAF
package. The DAOPHOT package was used to perform aperture photometry
of the GRB images. 
Standard star observation in one night is used to derive magnitudes of
reference stars in the vicinity of the GRB afterglow, and these
reference stars were used to perform photometry of the afterglow.  We
also made use of the Pan-STARRS1 $3\pi$ catalogs
\citep{ps1a,ps1b,ps1c} to calibrate our $g'$, $r'$, $i'$, $z'$, and
$Y$ band data.
As shown in Figure \ref{lc}, the light curves in the $g'$, $r'$,
$i'$, and $z'$ bands indicated the achromatic temporal break at
$\sim2.6\times10^{5}$ s. We successfully fitted the broken power-law
model to the $g'$, $r'$, $i'$, and $z'$ bands light curves.
Regarding the $g'$ band, we have obtained
$\alpha_{g1}=-1.01\pm0.06$,
$\alpha_{g2}=-2.84\pm0.16$, and 
$t_{bg}=(2.58\pm0.15)\times10^{5}$ s;
regarding the $r'$ band,
$\alpha_{r1}=-0.96\pm0.01$,
$\alpha_{r2}=-2.64\pm0.09$, and
$t_{br}=(2.58\pm0.07)\times10^{5}$ s;
regarding the $i'$ band,
$\alpha_{i1}=-0.88\pm0.02$,
$\alpha_{i2}=-2.48\pm0.05$, and
$t_{bi}=(2.51\pm0.10)\times10^{5}$ s;
regarding the $z'$ band,
$\alpha_{z1}=-0.89\pm0.01$,
$\alpha_{z2}=-2.64\pm0.10$, and
$t_{bz}=(2.63\pm0.06)\times10^{5}$ s;
and regarding the $y'$ band, we fitted the light curve by using the
simple power-law model, because the $y'$ band observations covered
only before the temporal break. The light curve was fitted with
the model and we obtained $\alpha_{y}=-0.87\pm0.03$.
The decay indices before and after the break are $\sim-1$ and
$\sim-2$, respectively, which is highly consistent with typical well-observed
long GRB optical afterglows. 
In Figure \ref{lcsed}, we plot the spectral flux
distribution with the submm and X-ray data. We fitted the optical data
alone using a power-law function and obtained $\beta = -1.44\pm0.10$,
$-1.11\pm0.09$, and $-1.18\pm0.17$ at $t=6.42\times10^{4}$ s,
$1.38\times10^{5}$ s, and $3.10\times10^{5}$ s, respectively.
To remove the effects of the Galactic interstellar extinction, we used
the reddening map by \citet{schlafly}.

%%%%% SMA analysis %%%%%%%%
The raw data of the SMA observations were calibrated using the MIR and
MIRIAD packages and images were made with the natural weighting.
%
%The resulting synthesized beam sizes and the achieved rms noise levels
%are summarized in Table 1.
%
Regarding the first night of observation, we split the data into three
periods to describe the temporal evolution of submm afterglow. 
Table \ref{tbl-2} summarizes each observation period and flux density
measurements (upper part).  Because of adverse weather conditions
during the first $1.08\times10^{4}$ s, only the data recorded after 13:00 on
March 26 UT were used for the scientific analysis. In the final period,
we constrained the 3$\sigma$ upper limit.
%faint afterglow was marginally detected at the 2.4$\sigma$ confident
%level. In the light curve shown in Figure \ref{lcsed} (left), we
%plotted the 3$\sigma$ limit.
%
With our SMA follow-ups, we successfully monitored the afterglow from
$4.32\times10^{4}$ to $3.46\times10^{5}$ s as shown in Figure
\ref{lc}.  The submm afterglow exhibited a flat evolution
with slight brightening between the first and second periods. To
describe the temporal evolution, we fitted the submm data with the
single power-law function and obtained $\alpha_{\rm
  submm}=-0.33\pm0.08$, which was considerably flatter than those of
the X-ray and optical.

\section{Discussion}

\subsection{Prompt Emission and Energetics Relations}

\begin{figure}
\plotone{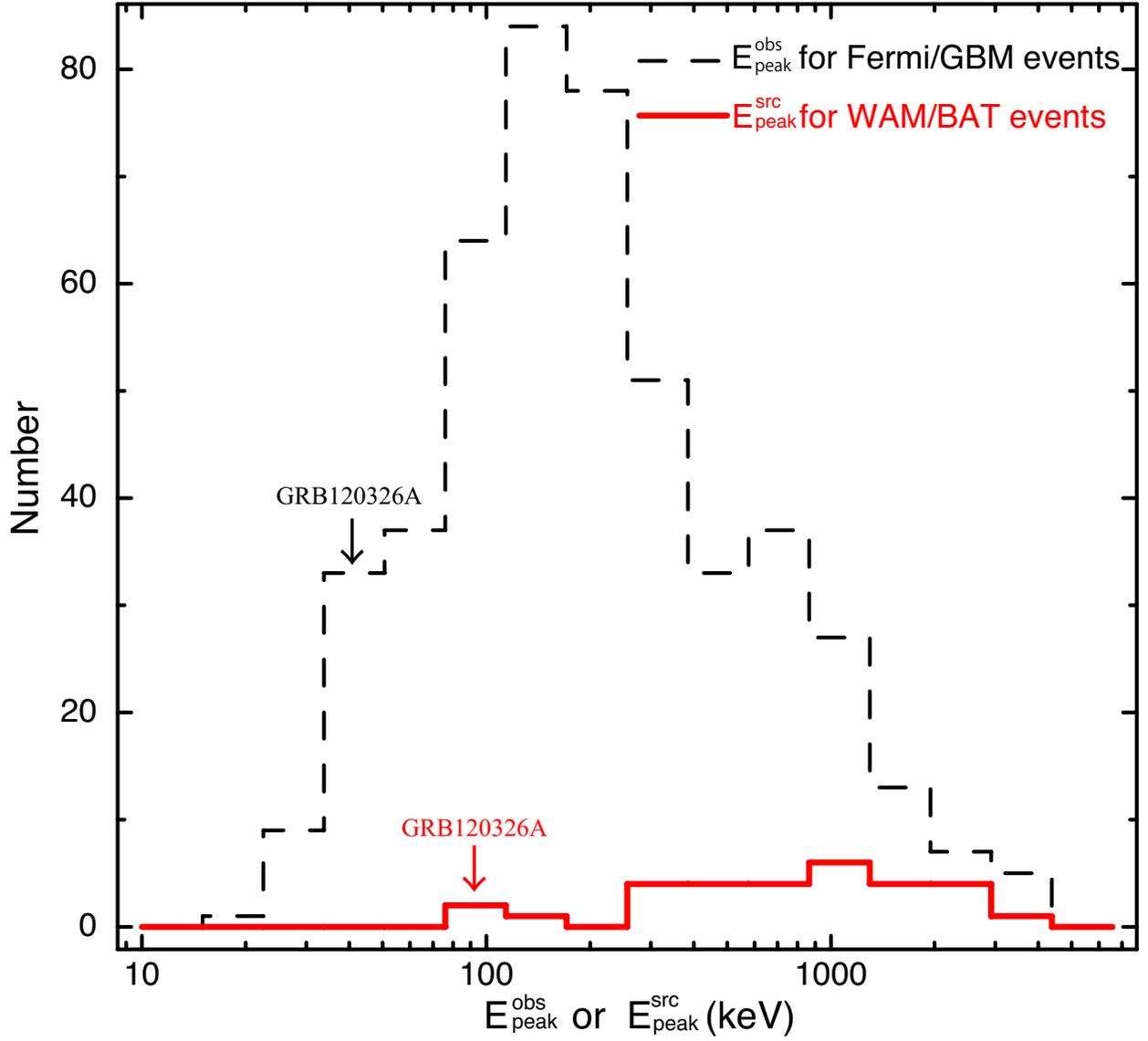}
\caption{The distribution of $E^{\rm src}_{\rm peak}$ for BAT/WAM events and $E^{\rm obs}_{\rm peak}$ for \fermi/GBM events. GRB120326A is one of the lowest $E^{\rm src}_{\rm peak}$ events among the BAT/WAM sample. $E^{\rm obs}_{\rm peak}$ also shows the same trend with large \fermi/GBM samples; although the measurements was analyzed without redshift correction. \label{epdist}}
\end{figure}

The joint fitting of the \Swift/BAT and \Suzaku/WAM suitably
constrained the spectral parameters of the prompt emission of
GRB120326A that are critical to characterize event. As shown in Figure
\ref{epdist}, the spectral peak energy in the source frame $E^{\rm
  src}_{peak}$ is one of the lowest events among the sample of the
joint \Swift/BAT-\Suzaku/WAM analysis. 
The trend is similar to that of the $\nu F \nu$ spectral peak energy
at the observer frame $E^{obs}_{peak}$ in comparison to a larger set
of $E^{obs}_{peak}$ values of 479 GRBs drawn from the \fermi/GBM
catalog \citet{gbmcat}; although, the \fermi/GBM measurements do not 
represent $E^{\rm src}_{peak}$ due to the lack of redshift information.
By comparing with the {\it HETE-2} sample \citep{hete}, GRB120326A
can be categorized as X-ray rich GRBs.  Using the definition with
\Swift/BAT data \citep{sakamoto}, we confirm that GRB120326A with
$\sim0.74$ fluence ratio in the 25-50 keV and 50-100 keV bands falls
into the X-ray rich GRB family.

\begin{figure}
\plotone{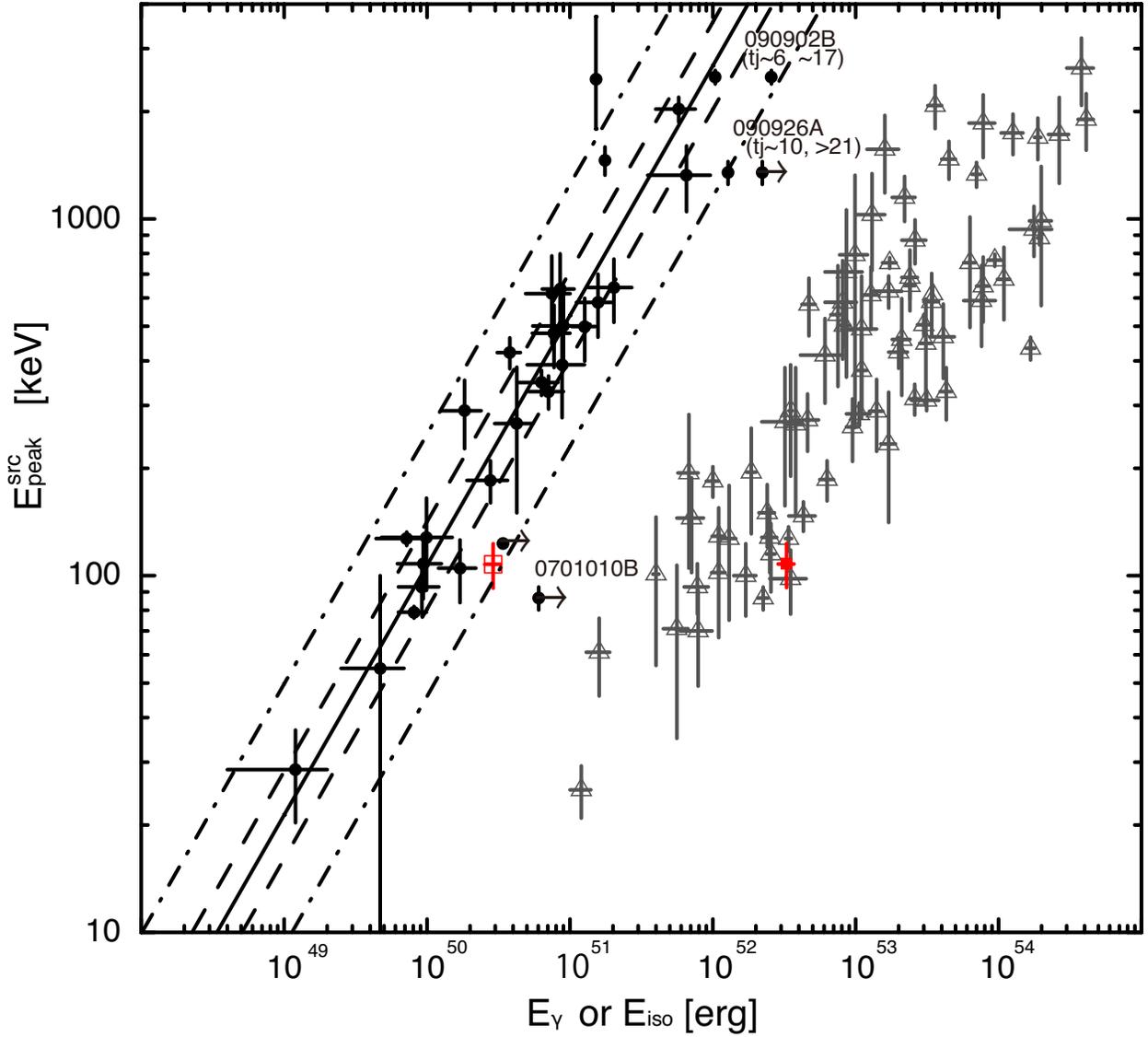}
\caption{The $E^{\rm src}_{\rm peak} - E_{\rm iso}$ relation in \citet{amati} (open triangle) and the $E^{\rm src}_{\rm peak}-E_{\rm \gamma}$ relation corrected for a homogeneous circumburst medium (filled circle). GRB120326A marked with red box points obeyed both relations.  The solid line indicates the best fit correlation  derived by \citet{g07}. The dashed and dash-dotted lines indicate the 1$\sigma$ and 3$\sigma$ scatter of the correlation, respectively.\label{relation}}
\end{figure}

The abundance of the multi-color optical light curves for estimating
the jet break time suggests that GRB120326A is a favorable target for
evaluating ${E^{\rm src}_{\rm peak} - E_{\rm iso}}$\citep{amati} and
${E^{\rm src}_{\rm peak}-E_{\rm \gamma}}$\citep{g07} relations.  Here,
${E^{\rm src}_{\rm peak}-E_{\rm \gamma}}$ is the correlation between
the intrinsic spectrum peak energy ${E^{\rm src}_{\rm peak}}$ and the
jet collimation-corrected energy in the prompt phase, ${E_{\rm
    \gamma}}$.
The closure relation of the observed optical temporal decay and the
spectral indices \citep[e.g.,][]{sari,bing} indicates that, of all
the $p>2$ options, the optical results are consistent with both the fast and
the slow cooling and both the wind and the ISM medium, as long as
$\nu_{m,FS}$ and $\nu_{c,FS}$ lie below the optical band.  Here,
$\nu_{m,FS}$ and $\nu_{c,FS}$ are the characteristic synchrotron
frequency and the cooling frequency based on the standard forward
shock synchrotron model.
Thus, the jet opening angle and the jet corrected energy are estimated
using $t_{br}$ as $6^{\circ}.69\pm0^{\circ}.16$ and
$(2.17\pm0.10)\times10^{50}$ erg by assuming the circumburst density
$n=$1.0 ${\rm cm^{-3}}$ and the energy conversion efficiency $\eta_{\gamma}=0.2$,
To covert the measured jet break time $t_{br}$ to the jet opening
angle, we used the formulation of \citet{sari} and \citet{frail}.
As shown in Figure \ref{relation}, GRB120326A obeys the \amati
and \ghil relations within $3 \sigma$ confidence level.
Therefore, GRB120326A belongs to the typical long duration GRB
family, even with a low $E^{\rm src}_{peak}$.

\subsection{Does Classical Forward Shock Synchrotorn model work?}

Based on the closure relations, the observed temporal evolution and
spectral features of the optical afterglow are well consistent with those
of the forward shock synchrotron model.
The fact that $\nu_{m,FS}$ and $\nu_{c,FS}$ both lie below the optical
band, implies that within the standard synchrotron model, X-ray
afterglow therefore lie in the same spectral regime as the optical
emission, and that the standard model therefore predicts the same
temporal and spectral shape for X-rays as for the optical.

However, the observed X-ray light curve shows a significant deviation
from the predicted behavior of the standard model and appears to
require an additional component.
Using the testing method of the forward shock model using the decay
index ($\alpha{\rm o}-\alpha_{X}$) relation between the optical and
X-ray\citep{normal}, we find that GRB120326A is a clear outlier
($\alpha{\rm o}-\alpha_{X}=-1.45\pm0.10$) and the origin of the X-ray
afterglow could differ from that of the optical.
For conducting a more rigorous analysis, we selected the normal
optical decay phase as from $3.2\times10^{4}$ s to $2.1\times10^{5}$
s. The optical light curve in this phase is well fitted with a simple
power law function with an index of $-0.96\pm0.01$.  For the time range,
we forced to fit the X-ray light curve with the simple power law
function and obtained the decay index of $-1.54\pm0.14$. Hence, this
event remained outlier ($\alpha{\rm o}-\alpha_{X}=-0.58\pm0.14$)
  and the X-ray emission could have a extra component such as
  X-ray flare on the forward shock synchrotron emission.

  To check the excess in the X-ray light curve, we shifted optical
  light curve to the X-ray band with the factor
  $(\nu_{X}/\nu_{opt})^{-p/2}$ by assuming the X-ray and optical lie
  on the same segment of the synchrotron radiation. We used
  $p=2.48$ with $3\sigma$ error of 0.15 estimated from the optical observations and the
  closure relation.  As shown in Figure \ref{lc}, the shifted light
  curve shows a significant gap with that of observed X-ray and the
  gap is rather smaller after $\sim3\times10^{5}$ s. One of the main
  reasons of the gap is no consideration of the intrinsic extinction
  for the optical component at the burst site due to lack of spectral
  coverage in the optical observations. 
  If the main X-ray component after $3\times10^{5}$ is originated from
  the same segment of forward shock synchrotron radiation with optical
  (as sharing the similr decay index with those of optical),
  $A_{V}=0.3\sim0.5$ mag with the SMC extinction curve is required to fill
  up the smaller gap after the $3\times10^{5}$ s.  The extinction
  value of $A_{V}=0.3\sim0.5$ mag is rather larger comparing with other
  majority of optically bright events ($A_{V}<0.25$ mag)
  \citep{kann06,kann10}. We also added the shifted optical light curve
  with $p=2$ that provides the upper limit of the forward shock
  synchrotron radiation under the $p>2$ condition (Figure
  \ref{lc}. Although these imply that adjustment of $p$ and
  introducing extinction may be the solution to explain the gap after
  $3\times10^{5}$, there remains a significant X-ray excess between
  $2\times10^{4}$ and $3\times10^{5}$ s.

  　The forward shock synchrotorn model based on the optical afterglow
  is also impossible to explain the submm emission.
  Based on the closure relation, the slow decay observed in
  the submm light curve requirs $\nu_{c,FS}<\nu_{submm}<\nu_{m,FS}$
  ({\it i.e}. fast cooling condition for both ISM and wind
  types). Hence, $\nu_{c,FS}<\nu_{submm}<\nu_{m,FS}<\nu_{opt}$ is
  required to satisfy the closure relation for the submm and optical
  afterglows all together.
  In addition to the closure relation, the observed flux densities
  (both in the submm and optical bands) at $6.42\times10^{4}$ s and
  the temporal decay index in the optical bands tightly constrain the
  range of characterized frequencies as $\nu_{c,FS}<2.3\times 10^{11}$
  Hz and $\nu_{m,FS}\sim 3\times 10^{14}$ Hz to make the fast cooling
  condition. With these conditions, we find that even a drastic case
  (e.g. $\epsilon_{B}\sim1$ and $\epsilon_{e}\sim1$ with very high
  density of $n>$ several $\times 10^{3}$ cm$^{\rm -3}$) cannot meet
  the condition and that the origin of submm component also could differ
  from that of the optical. Thus, additional radiation is required to
  explain X-ray and submm emissions all together.

%% Numerical modeling with boxfit.

\subsection{Foward Shock Synchrotron modeling}

To describe the entire spectral energy distribution, we performed
modeling for the optical light curves and spectra by using the boxfit
code \citep{boxfit} that involved two-dimensional relativistic
hydrodynamical jet simulations to determine the burst explosion and
the synchrotron radiation parameters with a homogeneous circumburst
medium. Hence, hereafter, we only consider the ISM condition and do
not verify whether ISM or wind condition is favorable.
This code also performs data fitting with the downhill simplex method
combined with simulated annealing.
Based on the observational results, we fixed as
$\theta_{jet}=6^{\circ}.7$ and a power-law electron spectrum of slope
$p=2.5$. The observing angle was also fixed as $\theta_{obs}=0$.
By using only the optical data with the code, we determined the optimal
modeling parameters to describe the optical light curves as $E=3.9\times
10^{52}$ erg, $n=1.0$ cm$^{\rm -3}$, $\epsilon_{B}=1.0\times 10^{-3}$,
and $\epsilon_{e}=6.9\times 10^{-1}$.
To adjust the model function, we also set the jet opening angle as a
free parameter and then $\theta_{jet}=8.1^{\circ}$ provided favorable
agreement ($\chi^{2}/\nu=3.9$ for $\nu=94$) with observing the light
curve. The solid line in Figure \ref{lc} indicates the best
model function for the $r'$ band light curve, which also well agreed
with the analytical model function.
This opening angle adjustment is reasonable because
$\theta_{jet}=6^{\circ}.7$ was estimated by considering the conical
jet. Additional note is that the post jet-break closure relation is no
longer valid under the consideration of detailed spreading of the jet
\citep{detailjet}. However, we are not concerned about this, since we
were able to achieve the good fit with the simulation-based fit
models.  With the adjusted jet opening angle, GRB120326A still obeyed
the \ghil relation.
We also attempted to determine the optimal solution by using optical
and submm data with the code. However, no sufficient solution describe
the temporal evolution of submm and optical afterglow all together was
determined by the $\chi^{2}$ evaluation with $2.3\times10^{4}$
iterations. All of trials provided reduced $\chi^{2}$ greater than 13.
This is consistent with the forward shock testings described above.
We generated the forward shock synchrotron model spectrum by using the
box fit with the best modeling parameters for the light curve. Figure
\ref{lcsed} shows the SED at $6.42\times10^{4}$ s after the
burst. The spectrum in the X-ray and submm exhibited substantial
excesses from the best model function and indicated that the afterglow
spectrum required additional radiation components.  This
interpretation is also consistent with the result of the
$\alpha{o}-\alpha_{X}$ relation and the shifted optical light curve to
the X-ray band with factor $(\nu_{X}/\nu_{opt})^{-p/2}$.

\subsection{Reverse Shock and Synchrotron  Self-Compton}

A solution that explains the X-ray excess and the different origin of
the submm emission is to introduce synchrotron self-inverse Compton
radiation from reverse shock.
This is one of the most feasible method to deal two notable observed
properties all together.
Assuming the deceleration time is near the X-ray light curve peak at
$t_{bX}\sim5.2\times10^{4}$ s, the initial Lorentz factor $\Gamma_{0}$
is estimated as $\sim16$, which is consistent with the thin-shell case
($\Gamma_{c}\sim365$). Here, $\Gamma_{c}$ is the critical Lorentz
factor that distinguishes thin shell models where the reverse shock
remains Newtonian, from thick shell models \citep{sari95, ssc}. This
lower $\Gamma_{0}$ might be associated with the low $E^{src}_{peak}$
property. 
It might also originate from the cocoon fireball as part of
two-component jet in the collapsar framework
\citep[e.g.,][]{cocoon}. In this case, thermal radiation is expected
to arise in the optical light curves
\citep{kashiyama,nakauchi}. However, the observed optical light curves
show no excess in the late phase.

Using the estimated parameters described in above, we
calculated the model function for synchrotron self-inverse Compton
radiation from reverse shock under the thin-shell condition described
in \citet{ssc}.
For this calculation, we assumed $\epsilon_{B,RS}\sim5\times10^{-3}$, 
and peak flux densities of reverse and
forward shocks as $F_{max,RS}\sim5.5$ mJy and $F_{max, FS}\sim 0.2$ mJy,
respectively.
Figure \ref{lcsed} shows the calculated spectrum for the reverse
shock and inverse Compton components at $6.42\times10^{4}$ s 
with the obtained key parameters of
$\nu^{Sync}_{m,RS}\sim4\times10^{11} {\rm Hz}$,
$\nu^{Sync}_{c,RS}\sim7\times10^{12} {\rm Hz}$,
$\nu^{IC}_{m,RS}\sim1\times10^{17} {\rm Hz}$, and
$\nu^{IC}_{c,RS}\sim4\times10^{19} {\rm Hz}$.
Although the observed X-ray flux was slightly brighter than calculated
self-inverse Compton component, the total spectrum including forward shock was
sufficiently described the overall properties of the afterglow.
Because $ \nu_{obs} < \nu^{Sync}_{m,RS}$, the expected decay index of
the reverse shock component in the observed submm band was
$\sim-0.46$, which was consistent with the slow temporal evolution of
the submm afterglow ($\alpha_{submm}=-0.33$). The relatively shallower
evolution than that of expected also implies the smooth transition of
$\nu^{Sync}_{m,RS}$ in the observing band, unlike the sharp break in
Figure \ref{lc}. This could be same with other spectrum
break such as non-existence of sharp cooling break in afterglow
spectrum \citep[e.g.,][]{granot02,van09,curran}.
The observed X-ray decay ($\alpha_{X}\sim-2.4$) and spectrum
($\beta_{X}\sim-0.96$) indices were also basically consistent with the
expected values ($\alpha_{X}\sim$2.8 and $\beta_{X}\sim-0.75$) for
$\nu^{IC}_{m,RS} < \nu_{obs} < \nu^{IC}_{c,RS}$.

\section{Summary}

We conducted multi-wavelength observations of a typical long duration
GRB 120326A, including rapid observations using SMA. Our SMA
observation successfully made the fastest afterglow detection among
seven submm afterglows at 230 GHz, and monitored from
$4.32\times10^{4}$ to $3.46\times10^{5}$ s.  The submm afterglow showed
considerably slower temporal evolution ($\alpha_{submm}=-0.33\pm0.08$)
which is unlikely to be explained by the forward shock synchrotron
model.
Based on our dense optical observations, we presented the optical
afterglows were well fitted by the broken power-law model and the
forward shock synchrotron model is feasible to explain the properties.
With the boxfit code, we also found the reasonable model function
within the forward shock synchrotron model under the assumption
of ISM circumburst medium.
Using the simple testing method of the forward shock model with 
temporal decay indices of optical and X-ray afterglows, we found the
origin of the X-ray afterglow could differ from that of optical.
Our joint spectrum fitting for prompt emission using \Swift/BAT and
\Suzaku/WAM also characterized the event and find that current event
obey the \amati and \ghil relations with in $3 \sigma$ confidence level.

Based on the detection and the slow decay of the afterglow in submm,
we introduced the synchrotron self-inverse Compton radiation from
reverse shock and find that this is a plausible idea to explain the
diversity.
This successful modeling could benefit other GRBs. Similar to
GRB120326A, numerous events exhibited no apparent jet breaks in X-ray
band and different temporal evolutions between the X-ray and
optical. These observational properties imply that additional
component such as reverse shock and its synchrotron self-inverse
Compton radiation make the different temporal evolution and hide
obvious jet break in X-ray. Because of a lack of submm observations
for these samples, interpretation from the same picture for these
events was difficult. Thus, further rapid follow-ups and continuous
monitoring with submm instruments such as SMA and Atacama Large
Millimeter/submillimeter Array (ALMA) will enable systematic testing
of the reverse shock and self-inverse Compton radiation.
 
\acknowledgments

We would like to thank Glen Petitpas for various arrangements on the
SMA observations and Shiho Kobayashi for useful comments. We would
also like to thank all of staff at the Lulin observatory. This work is
partly supported by the Ministry of Education and the National Science
Council of Taiwan grants NSC 100-2112-M-008-007-MY3(YU),
99-2112-M-002-002-MY3(KYH).
MI, JWK, MJ, and SJ acknowledge the support from the National Research
Foundation of Korea (NRF) grant, No. 2008-0060544, funded by the
Korea government (MSIP).
This work made use of data supplied by the UK Swift Science Data
Centre at the University of Leicester.
This paper includes data taken at the McDonald Observatory of the University
of Texas at Austin.
The PS1 Surveys have been made possible through contributions of the
Institute for Astronomy, the University of Hawaii, the Pan-STARRS
Project Office, the Max-Planck Society and its participating
institutes, the Max Planck Institute for Astronomy, Heidelberg and the
Max Planck Institute for Extraterrestrial Physics, Garching, The Johns
Hopkins University, Durham University, the University of Edinburgh,
Queen's University Belfast, the Harvard-Smithsonian Center for
Astrophysics, and the Las Cumbres Observatory Global Telescope
Network, Incorporated, the National Central University of Taiwan,
and the National Aeronautics and Space Administration under Grant
No. NNX08AR22G issued through the Planetary Science Division of the
NASA Science Mission Directorate.

\clearpage

\begin{deluxetable}{crccc}
\tablewidth{0pt} 	      	
\tabletypesize{\scriptsize} 
\tablecaption{Log of optical observations.\label{tbl-1}}
\tablehead{\colhead{Instruments} & \colhead{T-T$_{0}$ (s)} & \colhead{Filter} & \colhead{Exposure (s)} & \colhead{Flux density (mJy)}} 
\startdata
    CQUEAN &      35818 & $g$ &                  120 & $(5.031\pm1.114)\times10^{-2}$ \\
    CQUEAN &      36428 & $g$ &                  120 & $(5.843\pm0.364)\times10^{-2}$ \\
       LOT &      62495 & $g$ &                  300 & $(2.751\pm0.130)\times10^{-2}$ \\
       LOT &      66550 & $g$ &                  300 & $(2.781\pm0.132)\times10^{-2}$ \\
    CQUEAN &     118878 & $g$ &                  120 & $(1.549\pm0.072)\times10^{-2}$ \\
    CQUEAN &     203667 & $g$ &                  120 & $(9.670\pm1.290)\times10^{-3}$ \\
    CQUEAN &     203788 & $g$ &                  120 & $(7.200\pm1.310)\times10^{-3}$ \\
    CQUEAN &     203909 & $g$ &                  120 & $(9.820\pm1.150)\times10^{-3}$ \\
    CQUEAN &     210775 & $g$ &                  120 & $(8.840\pm2.050)\times10^{-3}$ \\
    CQUEAN &     210896 & $g$ &                  120 & $(9.580\pm2.540)\times10^{-3}$ \\
    CQUEAN &     211018 & $g$ &                  120 & $(1.460\pm0.271)\times10^{-2}$ \\
    CQUEAN &     291923 & $g$ &         120$\times$5 & $(5.800\pm1.330)\times10^{-3}$ \\
    CQUEAN &     382503 & $g$ &         300$\times$3 & $(2.600\pm0.320)\times10^{-3}$ \\
    CQUEAN &     468068 & $g$ &         300$\times$3 & $(1.550\pm0.420)\times10^{-3}$ \\
      LOAO &     119527 & $R$ &                  300 & $(7.517\pm0.848)\times10^{-2}$ \\
      LOAO &     119845 & $R$ &                  300 & $(6.792\pm0.654)\times10^{-2}$ \\
      LOAO &     120163 & $R$ &                  300 & $(8.395\pm0.596)\times10^{-2}$ \\
      LOAO &     120532 & $R$ &                  300 & $(1.019\pm0.055)\times10^{-1}$ \\
      LOAO &     120882 & $R$ &                  300 & $(1.096\pm0.059)\times10^{-1}$ \\
      LOAO &     121197 & $R$ &                  300 & $(8.395\pm0.596)\times10^{-2}$ \\
      LOAO &     206539 & $R$ &                  300 & $(4.787\pm0.579)\times10^{-2}$ \\
      LOAO &     206865 & $R$ &                  300 & $(4.365\pm0.633)\times10^{-2}$ \\
      LOAO &     207181 & $R$ &                  300 & $(5.105\pm0.740)\times10^{-2}$ \\
      LOAO &     207498 & $R$ &                  300 & $(3.565\pm0.373)\times10^{-2}$ \\
      LOAO &     301087 & $R$ &                  900 & $(2.148\pm0.242)\times10^{-2}$ \\
      LOAO &     300315 & $R$ &                  900 & $(2.128\pm0.275)\times10^{-2}$ \\
      LOAO &     303167 & $R$ &                 1200 & $(2.884\pm0.349)\times10^{-2}$ \\
      LOAO &     589749 & $R$ &                 2700 & $(5.750\pm0.880)\times10^{-3}$ \\
    CQUEAN &      31732 & $r$ &                  300 & $(1.734\pm0.022)\times10^{-1}$ \\
    CQUEAN &      33129 & $r$ &                  300 & $(1.636\pm0.008)\times10^{-1}$ \\
    CQUEAN &      34399 & $r$ &                  300 & $(1.510\pm0.027)\times10^{-1}$ \\
    CQUEAN &      36739 & $r$ &                  300 & $(1.478\pm0.010)\times10^{-1}$ \\
       LOT &      63503 & $r$ &                  300 & $(8.434\pm0.177)\times10^{-2}$ \\
       LOT &      67564 & $r$ &                  300 & $(8.591\pm0.165)\times10^{-2}$ \\
    CQUEAN &     117385 & $r$ &                  300 & $(4.955\pm0.036)\times10^{-2}$ \\
    CQUEAN &     202218 & $r$ &                  120 & $(1.812\pm0.207)\times10^{-2}$ \\
    CQUEAN &     202340 & $r$ &                  120 & $(2.526\pm0.175)\times10^{-2}$ \\
    CQUEAN &     202461 & $r$ &                  120 & $(2.544\pm0.141)\times10^{-2}$ \\
    CQUEAN &     209219 & $r$ &                  120 & $(2.586\pm0.236)\times10^{-2}$ \\
    CQUEAN &     209633 & $r$ &                  120 & $(2.720\pm0.116)\times10^{-2}$ \\
    CQUEAN &     209754 & $r$ &                  120 & $(2.496\pm0.119)\times10^{-2}$ \\
    CQUEAN &     286528 & $r$ &                  120 & $(1.715\pm0.075)\times10^{-2}$ \\
    CQUEAN &     286650 & $r$ &                  120 & $(1.712\pm0.067)\times10^{-2}$ \\
    CQUEAN &     286771 & $r$ &                  120 & $(1.603\pm0.065)\times10^{-2}$ \\
    CQUEAN &     295768 & $r$ &                  120 & $(2.072\pm0.291)\times10^{-2}$ \\
    CQUEAN &     381506 & $r$ &                  180 & $(8.050\pm0.380)\times10^{-3}$ \\
    CQUEAN &     381688 & $r$ &                  180 & $(7.040\pm0.390)\times10^{-3}$ \\
    CQUEAN &     381869 & $r$ &                  180 & $(8.500\pm0.380)\times10^{-3}$ \\
    CQUEAN &     466726 & $r$ &                  300 & $(5.300\pm0.710)\times10^{-3}$ \\
    CQUEAN &     467028 & $r$ &                  300 & $(3.770\pm0.670)\times10^{-3}$ \\
    CQUEAN &     467329 & $r$ &                  300 & $(4.860\pm0.550)\times10^{-3}$ \\
    CQUEAN &     552707 & $r$ &                  300 & $(2.890\pm0.270)\times10^{-3}$ \\
    CQUEAN &     553009 & $r$ &                  300 & $(2.930\pm0.290)\times10^{-3}$ \\
    CQUEAN &     553310 & $r$ &                  300 & $(3.070\pm0.300)\times10^{-3}$ \\
    CQUEAN &     639867 & $r$ &         300$\times$6 & $(1.970\pm0.300)\times10^{-3}$ \\
    CQUEAN &      32074 & $i$ &                  300 & $(3.621\pm0.010)\times10^{-1}$ \\
    CQUEAN &      33448 & $i$ &                  300 & $(3.701\pm0.005)\times10^{-1}$ \\
    CQUEAN &      34711 & $i$ &                  300 & $(3.803\pm0.023)\times10^{-1}$ \\
    CQUEAN &      37050 & $i$ &                  300 & $(3.311\pm0.039)\times10^{-1}$ \\
       LOT &      68583 & $i$ &                  300 & $(1.964\pm0.028)\times10^{-1}$ \\
       LOT &      64510 & $i$ &                  300 & $(1.982\pm0.021)\times10^{-1}$ \\
    CQUEAN &     117705 & $i$ &                  300 & $(1.257\pm0.003)\times10^{-1}$ \\
    CQUEAN &     202624 & $i$ &                  120 & $(6.634\pm0.084)\times10^{-2}$ \\
    CQUEAN &     202746 & $i$ &                  120 & $(6.164\pm0.082)\times10^{-2}$ \\
    CQUEAN &     202867 & $i$ &                  120 & $(7.226\pm0.084)\times10^{-2}$ \\
    CQUEAN &     209979 & $i$ &                  120 & $(5.834\pm0.089)\times10^{-2}$ \\
    CQUEAN &     210100 & $i$ &                  120 & $(6.346\pm0.102)\times10^{-2}$ \\
    CQUEAN &     210222 & $i$ &                  120 & $(6.188\pm0.103)\times10^{-2}$ \\
    CQUEAN &     286927 & $i$ &                  120 & $(4.194\pm0.069)\times10^{-2}$ \\
    CQUEAN &     287048 & $i$ &                  120 & $(4.292\pm0.099)\times10^{-2}$ \\
    CQUEAN &     287169 & $i$ &                  120 & $(4.360\pm0.071)\times10^{-2}$ \\
    CQUEAN &     287313 & $i$ &                  120 & $(4.186\pm0.063)\times10^{-2}$ \\
    CQUEAN &     295903 & $i$ &                  120 & $(4.150\pm0.235)\times10^{-2}$ \\
    CQUEAN &     296025 & $i$ &                  120 & $(5.662\pm0.943)\times10^{-2}$ \\
    CQUEAN &     380844 & $i$ &                  180 & $(2.118\pm0.036)\times10^{-2}$ \\
    CQUEAN &     381025 & $i$ &                  180 & $(2.092\pm0.038)\times10^{-2}$ \\
    CQUEAN &     381206 & $i$ &                  180 & $(2.278\pm0.036)\times10^{-2}$ \\
    CQUEAN &     465783 & $i$ &                  300 & $(1.256\pm0.056)\times10^{-2}$ \\
    CQUEAN &     466084 & $i$ &                  300 & $(1.178\pm0.060)\times10^{-2}$ \\
    CQUEAN &     466386 & $i$ &                  300 & $(1.390\pm0.057)\times10^{-2}$ \\
    CQUEAN &     553652 & $i$ &                  300 & $(7.760\pm0.280)\times10^{-3}$ \\
    CQUEAN &     553954 & $i$ &                  300 & $(8.820\pm0.280)\times10^{-3}$ \\
    CQUEAN &     554255 & $i$ &                  300 & $(8.200\pm0.270)\times10^{-3}$ \\
    CQUEAN &     648922 & $i$ &         300$\times$6 & $(4.880\pm0.290)\times10^{-3}$ \\
    CQUEAN &      32416 & $z$ &                  300 & $(9.325\pm0.016)\times10^{-1}$ \\
    CQUEAN &      33766 & $z$ &                  300 & $(9.250\pm0.021)\times10^{-1}$ \\
    CQUEAN &      35023 & $z$ &                  300 & $(9.018\pm0.073)\times10^{-1}$ \\
       LOT &      65534 & $z$ &                  300 & $(5.278\pm0.088)\times10^{-1}$ \\
       LOT &      69604 & $z$ &                  300 & $(4.358\pm0.085)\times10^{-1}$ \\
    CQUEAN &     118015 & $z$ &                  300 & $(3.025\pm0.006)\times10^{-1}$ \\
    CQUEAN &     203042 & $z$ &                  120 & $(1.639\pm0.020)\times10^{-1}$ \\
    CQUEAN &     203163 & $z$ &                  120 & $(1.970\pm0.032)\times10^{-1}$ \\
    CQUEAN &     203285 & $z$ &                  120 & $(1.660\pm0.024)\times10^{-1}$ \\
    CQUEAN &     210347 & $z$ &                  120 & $(1.744\pm0.018)\times10^{-1}$ \\
    CQUEAN &     210468 & $z$ &                  120 & $(1.650\pm0.018)\times10^{-1}$ \\
    CQUEAN &     210589 & $z$ &                  120 & $(1.744\pm0.019)\times10^{-1}$ \\
    CQUEAN &     287478 & $z$ &                  120 & $(1.113\pm0.017)\times10^{-1}$ \\
    CQUEAN &     287600 & $z$ &                  120 & $(1.112\pm0.017)\times10^{-1}$ \\
    CQUEAN &     287721 & $z$ &                  120 & $(1.147\pm0.016)\times10^{-1}$ \\
    CQUEAN &     380256 & $z$ &                  120 & $(5.236\pm0.079)\times10^{-2}$ \\
    CQUEAN &     380438 & $z$ &                  120 & $(5.696\pm0.083)\times10^{-2}$ \\
    CQUEAN &     380619 & $z$ &                  120 & $(5.524\pm0.078)\times10^{-2}$ \\
    CQUEAN &     464838 & $z$ &                  180 & $(2.944\pm0.141)\times10^{-2}$ \\
    CQUEAN &     465140 & $z$ &                  180 & $(2.680\pm0.103)\times10^{-2}$ \\
    CQUEAN &     465441 & $z$ &                  180 & $(2.204\pm0.100)\times10^{-2}$ \\
    CQUEAN &     554653 & $z$ &                  300 & $(1.784\pm0.060)\times10^{-2}$ \\
    CQUEAN &     554955 & $z$ &                  300 & $(2.248\pm0.062)\times10^{-2}$ \\
    CQUEAN &     555256 & $z$ &                  300 & $(2.120\pm0.064)\times10^{-2}$ \\
    CQUEAN &      32776 & $Y$ &                  300 & $1.585\pm0.004$ \\
    CQUEAN &     118338 & $Y$ &                  300 & $(5.005\pm0.025)\times10^{-1}$ \\
    CQUEAN &     205445 & $Y$ &         120$\times$6 & $(3.576\pm0.054)\times10^{-1}$ \\
    CQUEAN &     288007 & $Y$ &         120$\times$3 & $(2.672\pm0.053)\times10^{-1}$ \\
\enddata
\end{deluxetable}

\begin{deluxetable}{ccrccc}
\tablewidth{0pt} 	      	
\tabletypesize{\scriptsize} 
\tablecaption{Log of SMA observations.\label{tbl-2}}
\tablehead{\colhead{Instruments} & \colhead{Observing period (UT)} &  \colhead{T-T$_{0}$ (s)} &\colhead{Band (GHz)} & \colhead{Beam size} & \colhead{Flux density (mJy)}} 
\startdata
SMA & 2012-03-26 13:00-15:00 &   45571 & 230  & $1".74\times1".14$ & 2.84 $\pm$ 0.86\\
SMA & 2012-03-26 15:10-17:30 &   53971 & 230  & $1".69\times1".02$ & 3.56 $\pm$ 0.75\\
SMA & 2012-03-26 18:00-20:15 &   64171 & 230  & $2".03\times0".96$ & 3.36 $\pm$ 1.04\\
SMA & 2012-03-27 13:25-19:15 &  138480 & 230  & $1".62\times1".11$ & 2.38 $\pm$ 0.51\\
SMA & 2012-03-29 12:45-18:50 &  310290 & 230  & $0".62\times0".37$ & 1.76 $\pm$ 0.50\\
SMA & 2012-04-11 15:50-21:10 & 1377571 & 230  & $0".55\times0".42$ & $<1.44 (3\sigma)$\\ 
\enddata
\end{deluxetable}
\end{document}